\documentclass[12pt,a4paper]{article}

\usepackage{a4wide}
\usepackage{amsmath}
\usepackage{bm}
\usepackage{amssymb}
\usepackage{hyperref}
\usepackage{epic}

\newcommand{\ep}{\epsilon}

\newcommand{\cN}{{\cal N}}
\newcommand{\Op}{\mathcal{O}}
\newcommand{\HS}{S}
\newcommand{\M}{j}

\makeatletter
\def\mr@ignsp#1 {\ifx\:#1\@empty\else #1\expandafter\mr@ignsp\fi}%
\newcommand{\multiref}[1]{\begingroup
\xdef\mr@no@sparg{\expandafter\mr@ignsp#1 \: }%
\def\mr@comma{}%
\@for\mr@refs:=\mr@no@sparg\do{\mr@comma\def\mr@comma{,}\ref{\mr@refs}}%
\endgroup}
\makeatother

\begin{document}

\thispagestyle{empty}


\begin{center}
{\Large{\bf
Leading transcendental contribution\\[1mm]
to the four-loop universal anomalous dimension\\[3mm] in $\cN=4$ SYM
}
}
\vspace{15mm}

{\sc
V.~N.~Velizhanin}\\[5mm]

{\it Theoretical Physics Department\\
Petersburg Nuclear Physics Institute\\
Orlova Roscha, Gatchina\\
188300 St.~Petersburg, Russia}\\[5mm]

\textbf{Abstract}\\[2mm]
\end{center}

\noindent{We present the result of a {\it full direct} component calculation
for the term proportional to $\zeta_5$ in the four-loop universal anomalous dimension
of the twist-2 operators in $\cN =4$ supersymmetric Yang-Mills theory
for the first three even moments. We suggest the general form such
contribution for the arbitrary Lorentz spin of the twist-2 operators.
}
\newpage

\setcounter{page}{1}


Recently we have performed~\cite{Velizhanin:2008jd} a {\it full direct} component calculation of the four-loop
anomalous dimension of the Konishi operator~\cite{Konishi:1983hf} in the $\cN=4$ supersymmetric
Yang-Mills (SYM) theory. This operator is the most simple non-protected operator among
the BMN-operators~\cite{Berenstein:2002jq} for which all-loops asymptotic Bethe-ansatz (ABA)
was found~\cite{Minahan:2002ve,Bena:2003wd}
in the framework of 
AdS/CFT-correspondence~\cite{Maldacena:1997re}.

In spite of its simplicity the Konishi operator is the first operator, where
a finite size effect appear and the ABA is violated by the ``wrapping''
effect~\cite{SchaferNameki:2006ey,Kotikov:2007cy}.
Some times ago these wrapping corrections for the Konishi operator were computed by two
different ways from the both sides of AdS/CFT-correspondence. In the $\cN=4$ SYM
theory the calculations were performed in the superfield formalism and take
into account only diagrams did not included in the asymptotic Bethe-ansatz~\cite{Fiamberti:2007rj},
following Ref.~\cite{Sieg:2005kd}.
From superstring side~\cite{Bajnok:2008bm} the finite size effects
were take into account
using L\"{u}scher formulas~\cite{Luscher:1986pf}.
The results of both computations are in agreement after corrections from the
perturbative side.  Our result of the {\it full direct} calculation is the same and
confirms correctness all suggestions of the computations from
Refs.~\cite{Fiamberti:2007rj,Bajnok:2008bm}, including the correctness of the
asymptotic Bethe-ansatz up to the four loops.

Generalization of the wrapping effect on the higher spins/twists/loops looks very
interesting from the point of view a possibility to find a general expression
similar to the ABA for the anomalous dimensions of all operators.
Correctness of the ``modified'' Bethe-ansatz can be checked as in
Ref.~\cite{Kotikov:2007cy} with the analytical continuation of the universal
anomalous dimension for the twist-2 operators with arbitrary Lorentz spin
(arbitrary number of the covariant derivatives inside operator) with BFKL~\cite{Lipatov:1976zz}
predictions. However it is possible
to find a general expression for the
leading transcendental contribution to the universal anomalous dimension proportional to $\zeta_5$.
Following a maximal
transcendentality principle~\cite{Kotikov:2002ab}, which we used for the computation
of the three-loop universal anomalous dimension~\cite{Kotikov:2004er} from the QCD
results~\cite{Moch:2004pa}, one can easily established, that only three harmonic
sums $\HS_2$, $\HS_{-2}$ and $\HS_{1,1}$ can accompanied $\zeta_5$. To find coefficients in
the front of these harmonic sums, it is necessary to know the results for any three
moments of the universal anomalous dimension and one moment is already known: it is Konishi.
But below we will show, that a knowledge of two moments provides interesting information
about relations between these three harmonic sums.

To find another moment let's us start with
the ``QCD-like'' colour and $SU(4)$ singlet local Wilson twist-2
operators:
\begin{eqnarray}
\mathcal{O}_{\mu _{1},...,\mu _{\M}}^{g} &=&\hat{S}
G_{\rho \mu_{1}}^{a}{\mathcal D}_{\mu _{2}}
{\mathcal D}_{\mu _{3}}...{\mathcal D}_{\mu _{\M-1}}G_{\rho \mu _{\M}}^a\,,
\label{ggs}\\
\mathcal{O}_{\mu _{1},...,\mu _{\M}}^{\lambda } &=&\hat{S}
\bar{\lambda}_{i}^{a}\gamma _{\mu _{1}}
{\mathcal D}_{\mu _{2}}...{\mathcal D}_{\mu _{\M}}\lambda ^{a\;i}\,, \label{qqs}\\
\mathcal{O}_{\mu _{1},...,\mu _{\M}}^{\phi } &=&\hat{S}
\bar{\phi}_{r}^{a}{\mathcal D}_{\mu _{1}}
{\mathcal D}_{\mu _{2}}...{\mathcal D}_{\mu _{\M}}\phi _{r}^{a}\,,\label{phphs}
\end{eqnarray}
where ${\mathcal D}_{\mu_i }$ are covariant derivatives. The spinors $\lambda _{i}$ and
field tensor $G_{\rho \mu }$ describe gluinos and gluons, respectively, and
$\phi _{r}$ are the complex scalar fields appearing in the ${\mathcal N}=4$ SYM theory.
Indices $i=1,\cdots ,4$ and $r=1,\cdots ,3$ refer to $SU(4)$ and
$SO(6)\simeq SU(4)$ groups of inner symmetry, respectively. The symbol
$\hat{S}$ implies a symmetrization of each tensor in the Lorentz indices
$\mu_{1},...,\mu _{\M}$ and a subtraction of its traces.
These operators are mixed with each other under renormalization and the matrix
of the anomalous dimensions known up to two loops~\cite{LN4,Kotikov:2002ab,Kotikov:2003fb}.
The eigenvalues of this
matrix can be expressed trough one function, the universal anomalous
dimension $\gamma_{uni}$, with a shifted argument. The universal anomalous dimension poses
the maximal transcendentality property, the hypothesis about which was suggested in
Ref.~\cite{Kotikov:2002ab} and then was conformed in the next-to-leading order with the full direct
computations in Ref.~\cite{Kotikov:2003fb}.

In the leading order the diagonalization of the anomalous dimension matrix is
equivalent to the construction of three multiplicative renormalizable
operators~\cite{LN4,Kotikov:2002ab}:
\begin{eqnarray}
\Op^T_{\mu _{1},...,\mu _{\M}} & = &
\Op^g_{\mu _{1},...,\mu _{\M}}
+ \Op^\lambda_{\mu _{1},...,\mu _{\M}}
+ \Op^\phi_{\mu _{1},...,\mu _{\M}}\,,
\label{mrop1}\\[1mm]
\Op^\Sigma_{\mu _{1},...,\mu _{\M}} & = &
- 2(j-1)\Op^g_{\mu _{1},...,\mu _{\M}}
+ \Op^\lambda_{\mu _{1},...,\mu _{\M}}
+ \frac{2(j+1)}{3}\Op^\phi_{\mu _{1},...,\mu _{\M}}\,,
\label{mrop2}\\[1mm]
\Op^\Xi_{\mu _{1},...,\mu _{\M}} & = &
- \frac{j-1}{j+2}\Op^g_{\mu _{1},...,\mu _{\M}}
+ \Op^\lambda_{\mu _{1},...,\mu _{\M}}
- \frac{j+1}{j}\Op^\phi_{\mu _{1},...,\mu _{\M}}\,.
\label{mrop3}
\end{eqnarray}

In the next-to-leading order only the first operator serves the
multiplicative renormalizability. It is related with the breaking of the
conformal invariance if we consider more general conformal
operators~\cite{Makeenko:1980bh}.
Breaking of conformal invariance is controlled by the conformal Ward
identity~\cite{Mueller:1991gd} (see also~\cite{Mikhailov:1985cm}),
which allows obtain the results for the anomalous dimensions of the conformal
operators in $n^{th}$-loops order with additional $(n\!-\!1)$-loops
calculations~\cite{Belitsky:1998vj}.
But in the leading order the multiplicative renormalizability allows to find all
three eigenvalues of the anomalous dimension matrix from its three
elements, for example, from $\gamma_{g\lambda}$, $\gamma_{\phi\lambda}$ and
$\gamma_{\lambda\lambda}$ (see~\cite{LN4,Kotikov:2002ab}).
Namely, if we take the matrix elements for the operators $\Op^T$, $\Op^\Sigma$ and $\Op^\Xi$
sandwiched between fermion states we obtain the following expressions
for the anomalous dimensions of these operators through the anomalous dimensions
of the operators $\Op^g$, $\Op^\lambda$ and $\Op^\phi$ sandwiched between fermion states:
\begin{eqnarray}
\Gamma^{(0)}_{T}(j) & = &
  \gamma^{(0)}_{g\lambda}(j)
+ \gamma^{(0)}_{\lambda\lambda}(j)
+ \gamma^{(0)}_{\phi\lambda}(j)=\gamma_{uni}^{(0)}(j-2)\,,
\label{admrop1}\\[1mm]
\Gamma^{(0)}_{\Sigma}(j) & = &
- 2(j-1)\gamma^{(0)}_{g\lambda}(j)
+ \gamma^{(0)}_{\lambda\lambda}(j)
+ \frac{2(j+1)}{3} \gamma^{(0)}_{\phi\lambda}(j)=\gamma_{uni}^{(0)}(j)\,,
\label{admrop2}\\[1mm]
\Gamma^{(0)}_{\Xi}(j) & = &
- \frac{j-1}{j+2}\gamma^{(0)}_{g\lambda}(j)
+ \gamma^{(0)}_{\lambda\lambda}(j)
- \frac{j+1}{j} \gamma^{(0)}_{\phi\lambda}(j)=\gamma_{uni}^{(0)}(j+2)\,
\label{admrop3}
\end{eqnarray}
with
\begin{equation}
\gamma_{uni}(j)=\gamma_{uni}^{(0)}(j)\;g^2
+\gamma_{uni}^{(1)}(j)\;g^4
+\gamma_{uni}^{(2)}(j)\;g^6
+\gamma_{uni}^{(3)}(j)\;g^8
+...
\,,\qquad
g^2=\frac{g^2_{YM}N_c}{16\pi^2}\,.
\end{equation}
Substitute explicit expressions for the anomalous dimensions
$\gamma^{(0)}_{g\lambda}$, $\gamma^{(0)}_{\phi\lambda}$ and
$\gamma^{(0)}_{\lambda\lambda}$~\cite{LN4,Kotikov:2002ab}:
\begin{eqnarray}
\gamma^{(0)}_{g\lambda}(j) &=& -\frac{8}{j-1}+\frac{8}{j}-\frac{4}{j+1}\,,\\[1mm]
\gamma^{(0)}_{\lambda\lambda}(j)&=& 8\; \HS_1(j) - \frac{16}{j} + \frac{16}{j+1}\,,\\[1mm]
\gamma^{(0)}_{\phi\lambda}(j)&=&-\frac{12}{j+1} \,,\label{olad}
\end{eqnarray}
it is easily to verify, that in the leading order the universal anomalous dimension is expressed
through the most simple harmonic sum $\HS_1$~\cite{LN4,Kotikov:2002ab}:
\begin{eqnarray}
\gamma^{(0)}_{uni}(j) &=& 8\; \HS_1(j) \,, \qquad \HS_1(j)\ =\ \sum^j_{m=1} \frac{1}{m}\,.\label{oluad}
\end{eqnarray}
We see that in the leading order a knowledge of three elements
$\gamma^{(0)}_{g\lambda}$, $\gamma^{(0)}_{\phi\lambda}$ and
$\gamma^{(0)}_{\lambda\lambda}$ from the anomalous dimension matrix allows to find for given $j$ three
different values of the universal anomalous dimension $\gamma_{uni}^{(0)}$.

If we go to the four loop we can do the same as in the one loop for the
contribution to the universal anomalous dimension, which is
proportional to $\zeta_5$ because $\zeta_5$ appears only in the
four-loop diagrams and then there are no additional contributions neither
from the renormalizations or from the conformal Ward identities.

The most simple operators in Eqs.~(\ref{mrop1})-(\ref{mrop3}) are the operators with $j\!=\!2$,
i.e. the following operators:
\begin{eqnarray}
\Op^T_{\mu\nu} & = &
\Op^g_{\mu\nu}
+ \Op^\lambda_{\mu\nu}
+ \Op^\phi_{\mu\nu}\,,
\label{mrop1j2}\\[1mm]
\Op^\Sigma_{\mu\nu} & = &
- 2\,\Op^g_{\mu\nu}
+ \Op^\lambda_{\mu\nu}
+ 2\,\Op^\phi_{\mu\nu}\,,
\label{mrop2j2}\\[1mm]
\Op^\Xi_{\mu\nu} & = &
- \frac{1}{4}\Op^g_{\mu\nu}
+ \Op^\lambda_{\mu\nu}
- \frac{3}{2}\Op^\phi_{\mu\nu}\,.
\label{mrop3j2}
\end{eqnarray}
Note, that the coefficients in the front of the operators
$\Op^g_{\mu\nu}$, $\Op^\lambda_{\mu\nu}$ and $\Op^\phi_{\mu\nu}$ in Eqs.~(\ref{mrop1j2})-(\ref{mrop3j2})
are the same (up to common factor) as in the conformal operators $\Xi_{\mu\nu}$,
$\Sigma_{\mu\nu}$ and $T_{\mu\nu}$ from Ref.~\cite{Anselmi:1998ms}.
Operator $\Op^T_{\mu\nu}$ is the stress tensor. Its anomalous dimension is equal to
zero and corresponds to $\gamma_{uni}(j\!=\!2)$.
Operator $\Op^\Sigma_{\mu\nu}$ has the same anomalous dimension as the Konishi operator, which corresponds to
$\gamma_{uni}(j\!=\!4)$ and we
already know  its anomalous dimension up to the four loops.
Operator $\Op^\Xi_{\mu\nu}$ has the anomalous dimension, which
corresponds to the value of universal anomalous dimension $\gamma_{uni}(j)$ with $j\!=\!6$.

So, we need to calculate only the four-loop diagrams with insertions of
the operator $\Op^\Xi_{\mu\nu}$
and looking only for the pole with $\zeta_5$.
It is possible to do with our program BAMBA~\cite{BAMBA}, applying Laporta's algorithm~\cite{Laporta:2001dd}
(see also~\cite{Misiak:1994zw,Czakon:2004bu}), which we used for the calculation of the
four-loop Konishi~\cite{Velizhanin:2008jd}. For the present calculations we extended our
database for the scalar integrals\footnote{Results for integrals can be obtained under request.}
containing $\zeta_5$ with additional powers of the denominators
and numerators\footnote{For testing the well-developed MATHEMATICA package FIRE~\cite{Smirnov:2008iw}
we have performed the same calculations with FIRE and have found a full agreement with
our results}.

All calculations were performed with FORM~\cite{Vermaseren:2000nd},
using FORM package COLOR~\cite{vanRitbergen:1998pn} for evaluation of the color traces
and with the Feynman rules from Refs.~\cite{Gliozzi:1976qd}.
For the dealing with a huge number of diagrams we use a program DIANA~\cite{Tentyukov:1999is},
which call QGRAF~\cite{Nogueira:1991ex} to generate all diagrams.

Really we have computed the $\zeta_5$ contributions to the anomalous dimensions
$\gamma_{g\lambda}$, $\gamma_{\phi\lambda}$ and $\gamma_{\lambda\lambda}$
of the operators $\Op^g_{\mu\nu}$, $\Op^\lambda_{\mu\nu}$ and $\Op^\phi_{\mu\nu}$
sandwiched between the fermion states to have a possibility to combine its with
coefficients from Eqs.~(\ref{mrop1j2})-(\ref{mrop3j2}) for the additional check.
We have obtained the following results for the anomalous dimensions
of the operators $\Op^T_{\mu\nu}$, $\Op^\Sigma_{\mu\nu}$ and $\Op^\Xi_{\mu\nu}$
sandwiched between fermion states:
\begin{eqnarray}
\Gamma_{T_{\mu\nu}} & = &
0\,,
\label{mrop1j2ad}\\[5mm]
\Gamma_{\Sigma_{\mu\nu}} & = &
- 1440\ \zeta_5\ g^8\,,
\label{mrop2j2ad}\\[2mm]
\Gamma_{\Xi_{\mu\nu}} & = &
-\frac{25000}{9}\ \zeta_5\  g^8\,.
\label{mrop3j2ad}
\end{eqnarray}
The first two results coincide with the expected results. Third result is new and
can be used to find the coefficients in the ansatz for the term proportional to $\zeta_5$ in the
four-loop universal anomalous dimension $\hat\gamma_{uni}^{(3)}$.
In the fourth order of the perturbative theory a transcendentality level
of the universal anomalous dimension is equal to 7 and the transcendentality of $\zeta_5$ is 5. Then,
the harmonic sums in the ansatz should have the transcendentality 2 and we should exclude
the harmonic sums with index ``-1". There are three such harmonic sums in the canonical
form (see Ref.~\cite{Vermaseren:1998uu} about harmonic sums):
\begin{equation}
\HS_2(j)=\sum_{m=1}^j\frac{1}{m^2}\,,\qquad
\HS_{-2}(j)=\sum_{m=1}^j\frac{(-1)^m}{m^2}\,,\qquad
\HS_{1,1}(j)=\sum_{m=1}^j\frac{1}{m}\sum_{k=1}^m\frac{1}{k}\,.
\end{equation}
For the last sum there is a relation $2\HS_{1,1}=\HS_{1}^2+\HS_{2}$ and
it is suitable to use $\HS_{1}^2$ instead of $\HS_{1,1}$.
Using two results~(\ref{mrop2j2ad}) and~(\ref{mrop3j2ad})
for three coefficients in the ansatz we have found the following
expression for the $\zeta_5$-term in the universal anomalous dimension
\begin{equation}\label{HSZ5Res}
\hat\gamma_{uni}^{(3)}(\M)\ =\ -\,640\,\Big(\big(32\;\HS_{-2}(\M) + 21\;\HS_2(\M)\big)\,\big(1-x\big)
+\,x\,\HS_{1}^2(\M)
\Big)\,\zeta_5\,,
\end{equation}
with arbitrary rational number $x$.
Surprisingly, that for $x=1$ we get
\begin{equation}\label{HSZ5ResS1}
\hat\gamma_{uni}^{(3)}(\M)\ =\ -\,640\,\HS_{1}^2(\M)\,\zeta_5\,.
\end{equation}
We believe, that this expression is more natural
from the point of view
a possible incorporation into the ``modified'' Bethe-ansatz.
Moreover, let's us rewrite our result through the ``wrapping'' integral, i.e.
the unique four-loop master-integral proportional to $\zeta_5$\\[-1mm]
\begin{figure}[h]
\begin{picture}(40,45)
\put(100,30){\circle{40}}
\put(75,30){\line(1,0){50}}
\put(100,5){\line(0,1){50}}
\put(100,30){\circle*{3}}
\put(120,30){\circle*{3}}
\put(100,50){\circle*{3}}
\put(100,10){\circle*{3}}
\put(80,30){\circle*{3}}
\end{picture}
\end{figure}
\vspace{-21mm}\begin{equation}{\mathcal I}(4)=\frac{5}{\ep}\;\zeta_5\,,\qquad
\hat{\mathcal I}(4)=5\;\zeta_5\,,
\end{equation}\\[3mm]
where the argument of the master-integral is the number of loops and we obtain
\begin{equation}\label{HSZ5ResS1I4}
\hat\gamma_{uni}^{(3)}(\M)\ =\ -\,128\,\HS_{1}^2(\M)\,\hat{\mathcal I}(4)\,.
\end{equation}
It is natural to suggest, that such form of the leading transcendental contribution to the
universal anomalous dimension will be hold also in higher loops, so
\begin{equation}\label{HSZ5ResS1I}
\hat\gamma_{uni}^{(N_{l}-1)}(\M)\ =\ -\,128\,\HS_{1}^2(\M)\,\hat{\mathcal I}(N_{l})\,.
\end{equation}

For possible check we have evaluated numerically the five-loop ``wrapping'' integral with 
FIESTA~\cite{Smirnov:2008py} and have reconstructed its value suggesting its proportionality to $\zeta_7$
\begin{center}
\hspace*{-160mm}
\begin{picture}(40,50)
\put(100,30){\circle{40}}
\put(75,30){\line(1,0){25}}
\put(100,30){\line(1,-2){11}}
\put(100,30){\line(-1,-2){11}}
\put(100,30){\line(1,0){25}}
\put(100,30){\line(0,1){25}}
\put(100,30){\circle*{3}}
\put(120,30){\circle*{3}}
\put(100,50){\circle*{3}}
%
\put(91,12){\circle*{3}}
\put(109,12){\circle*{3}}
\put(80,30){\circle*{3}}
\end{picture}
\vspace*{-20mm}
\begin{equation}{\mathcal I}(5)=\frac{14}{\ep}\;\zeta_7\,,\qquad
\hat{\mathcal I}(5)=14\;\zeta_7\,.
\end{equation}\\[2mm]
\end{center}

Probably, a general form of the leading transcendental contribution to the
universal anomalous dimension Eq.~(\ref{HSZ5ResS1I}) hold not for $\hat\gamma_{uni}$ but for
the renormalization constant, which related with anomalous dimension as the derivative with
respect to coupling constant and then Eq.~(\ref{HSZ5ResS1I}) should be replaced by
\begin{equation}\label{HSZ5ResS1IL}
\hat\gamma_{uni}^{(N_{l}-1)}(\M)\ =\ -\,32\,N_l\,\HS_{1}^2(\M)\,\hat{\mathcal I}(N_{l})\,.
\end{equation}

Moreover, based on the criterion of simplicity we suggest the following ad hoc expression
for the term proportional
to the $\zeta_3$ in the four-loop universal anomalous dimension
coming from the ``wrapping'' effect with the addition to the result from the
asymptotic Bethe-ansatz (see Table 1 in Ref.~\cite{Kotikov:2007cy}):
\begin{equation}\label{HSZ3Res}
\hat\gamma_{uni}^{(3)}(\M)\ =\ 256\,
\Big(3\,\HS_4 - \big(\HS_{-4} - 2\,\HS_{-3, 1} - 2\,\HS_{-2, 2} + 4\,\HS_{-2, 1, 1}\big)\Big)
\zeta_3\,.
\end{equation}
This expression based on the fact, that the ``wrapping'' corrections did not modified
the large-$j$ behavior of the ABA and our experience of working with the harmonic sums.

In any case our result~(\ref{mrop3j2ad})
of the {\it full direct} four-loop calculations for the term proportional to $\zeta_5$
in the four-loop universal anomalous dimension $\gamma_{uni}^{(3)}(j)$ for the twist-2 operator with $j\!=\!6$
is new and can be used to check the similar result of the ``wrapping'' corrections.

 \subsection*{Acknowledgments}
We would like to thank L.N. Lipatov, A.I. Onishchenko, A.V. Smirnov, V.A. Smirnov and M. Staudacher
for useful discussions.
This work is supported by
RFBR grants 07-02-00902-a, RSGSS-3628.2008.2.
We thank the Max Planck Institute for Gravitational Physics
at Potsdam for hospitality while working on parts of this
project.

\end{document}